\documentstyle[epsfig,12pt]{article}

\title{\begin{flushright}
{\normalsize NUC-MINN-2002/10-T\\
November 2002 \\}
\end{flushright}
\vspace*{0.3in}
{\bf HIGH TEMPERATURE MATTER AND NEUTRINO SPECTRA FROM MICROSCOPIC BLACK 
HOLES}}
\author{{\bf R. G. Daghigh and J. I. Kapusta} \vspace*{0.1in}\\
 {\it School of Physics and Astronomy, University of Minnesota}\\
 {\it Minneapolis, MN 55455, USA}}

\date{}

\begin{document}

\maketitle

\begin{abstract}
The spectrum of neutrinos produced by the outflow of high temperature matter 
surrounding microscopic black holes is calculated for neutrino energies between 
one GeV and the Planck energy.  The results may be applicable for the last few 
hours and minutes of a microscopic black hole's lifetime.
\end{abstract}

\centerline{\small PACS: 04.70.Dy, 98.70.Sa, 95.85.Ry, 95.55.Vj}

\section{Introduction}

Hawking radiation from black holes \cite{Hawk} is of fundamental interest 
because it involves the interplay between quantum field theory and the strong 
field limit of general relativity.  It is of astrophysical interest if black 
holes exist with sufficiently small mass that they explode rather than accrete 
matter and radiation.  Since the Hawking temperature and black hole mass are 
related by $T_H = m_{\rm P}^2/8\pi M$, where $m_{\rm P} = G^{-1/2} = 1.22\times 
10^{19}$ GeV is the Planck mass with natural units $\hbar = c = k_{\rm B} = 1$, 
a present-day black hole will evaporate and eventually explode only if $T_H > 
2.7$ K (microwave background temperature).  This requires    
$M < 4.6\times 10^{22}$ kg which is approximately 1\% of the mass of the Earth.
More massive black holes are cooler and therefore will absorb more matter and
radiation than they emit, hence grow with time.  Taking into account emission
of gravitons, photons, and neutrinos a critical mass black hole today has a
Schwarszchild radius of 68 microns and a lifetime of $2\times10^{43}$ years.

There is uncertainty expressed in the literature about whether the radiation 
emitted by such small black holes is emitted independently, or whether there is 
sufficient interaction among the emitted quanta to form a partially thermalized 
and rapidly expanding fluid around the black hole.  Heckler has given arguments 
in favor of the latter for Hawking temperatures above 1 GeV \cite{Heckler}.  
Cline, Mostoslavsky and Servant \cite{cline} solved the Boltzmann equation in 
the relaxation-time approximation for such temperatures and found that 
significant particle scattering did occur, although not enough for perfect fluid 
flow.  One of us showed that a self-consistent description of an outgoing fluid 
just marginally kept in local equilibrium could be given, but it required the 
assumption of sufficient initial particle interaction, and viscosity played a 
crucial role \cite{me}.  That paper was followed by an extensive numerical 
analysis of the relativistic viscous fluid equations by us \cite{us}.  We also 
calculated the spectrum of high energy gamma rays expected during the last days, 
hours, and minutes of the black hole's lifetime.  We suggested that the most 
promising route for discovery of such microscopic black holes is to search for 
point sources emitting gamma rays of ever-increasing energy until suddenly the 
source shuts off.

In this paper, a follow up to our previous work, we focus on high energy 
neutrino emission from black holes with Hawking temperatures greater than 100 
GeV and corresponding masses less than 10$^8$ kg.  It is at these and higher 
temperatures that new physics will arise.  Such a study is especially important 
in the context of high energy neutrino detectors under construction or planned 
for the future.  Previous notable studies in this area have been carried out by 
MacGibbon and Webber \cite{nu4} and by Halzen, Keszthelyi and Zas \cite{nu5}, 
who calculated the instantaneous and time-integrated spectra of neutrinos 
arising from the decay of quark and gluon jets. 

The source of neutrinos in the viscous fluid picture is quite varied.  Neutrinos 
should stay in thermal equilibrium, along with all other elementary particles, 
when the local temperature is above 100 GeV.  The reason is that 
at energies above the electroweak scale of 100 GeV neutrinos should have 
interaction cross sections similar to those of all other particles.  Thus the 
neutrino-sphere, where the neutrinos decouple, ought to exist where the local 
temperature falls below 100 GeV.  The spectra of these direct neutrinos are 
calculated in section II.

Pions and muons remain in local thermal equilibrium down to temperatures on the 
order of 100 to 140 MeV, as shown in \cite{us}.  Neutrinos also come from decays
involving these particles.  The relevant processes are (i) a thermal pion decays 
into a muon and muon-neutrino, followed by the muon decay $\mu \rightarrow e 
\nu_e \nu_{\mu}$, and (ii) a thermal muon decays in the same way. 
The spectra of neutrinos arising from pion decay are calculated in section III 
while those arising from direct or indirect muon decay are calculated in section 
IV.

The spectra from all of these sources are compared graphically in section V.  
We also compare with the spectra of neutrinos emitted directly as Hawking 
radiation without any subsequent interactions.  The main result is that the 
time-integrated direct Hawking spectrum falls at high energy as $E^{-3}$ whereas 
the time-integrated neutrino spectrum coming from a fluid or from pion and muon 
decays all fall as $E^{-4}$.  Thus the fluid picture predicts more neutrinos at 
lower energies than the direct Hawking emission picture.  If a microscopic black 
hole is near enough the instantaneous spectrum could be measured, and its shape 
and magnitude would provide information on the number of degrees of freedom in 
the nature on mass scales exceeding 100 GeV.  Conclusions are presented in 
section VI.

Any details not provided here on the outflow of high temperature matter 
surrounding a black hole may be found in \cite{me,us}.

\section{Directly Emitted Neutrinos}

In this section we first review the emission of neutrinos by the Hawking 
mechanism unmodified by any rescattering.  Then we estimate the spectra of 
neutrinos which rescatter in the hot matter.  The last scattering surface should 
be represented approximately by that radius where the temperature has dropped to 
100 GeV.  The reason is that neutrinos with energies much higher than that have 
elastic and inelastic cross sections that are comparable to the cross sections 
of quarks, gluons, electrons, muons, and tau leptons.  Much below that energy 
the relevant cross sections are greatly suppressed by the mass of the exchanged 
vector bosons, the W and Z.   Furthermore, the electroweak symmetry is broken 
below temperatures of this order, making it natural to place the last scattering 
surface there.  A much better treatment would require the solution of transport 
equations for the neutrinos, an effort that is perhaps not yet justified.  All 
the formulas in this section refer to one flavor of neutrino.  Our current 
understanding is that there are three flavors, each available as a particle or 
antiparticle.  The sum total of all neutrinos would then be a factor of 6 larger 
than the formulas presented here.  

\subsection{Direct neutrinos}

The emission of neutrinos by the Hawking mechanism is usually calculated on the 
basis of detailed balance.  It involves a thermal flux of neutrinos incident on 
a black hole.  The Dirac equation is solved and the absorption coefficient is 
computed.  This involves numerical calculations \cite{nu1}-\cite{nu3}.
The number emitted per unit time per unit energy is given by
\begin{eqnarray}
\frac{d^2N_{\nu}^{\rm dir}}{dEdt} = \frac{\Gamma_{\nu}}{2\pi} \,
\frac{1}{\exp(E/T_H)+1} \,,
\end{eqnarray}
where $\Gamma_{\nu}$ is an energy-dependent absorption coefficient.  There is no 
simple analytic formula for it. For our purposes it is sufficient to parametrize 
the numerical results.  A fair representation is given by
\begin{eqnarray}
\Gamma_{\nu} = \frac{27}{64\pi^2} \frac {E^2}{T_H^2} \left( 0.075+\frac{0.925}
{\exp(5-aE/T_H)+1} \right),
\end{eqnarray}
and $a \approx 1.607$.  The exact expression has very small amplitude 
oscillations arising from the essentially black disk character of the black 
hole.  The parametrized form does not have these oscillations, but otherwise is 
accurate to within about 5\%.

If there are no new degrees of freedom present in nature, other than those 
already known, then the time dependence of the black hole mass and temperature 
are easily found.  The temperature is
\begin{equation}
T_H(t) = \frac{1}{8\pi} \left( \frac{-m_P^2}{3 \alpha t} \right)^{1/3} \, .
\end{equation}
Here the temperature is $T_H(t_0) = T_0$ at (negative) time $t_0$ and increases 
to infinity at time $t=0$.  The constant $\alpha$ is approximately 0.0043 for 
$T_H > 100$ GeV.  This relationship between the time and the temperature allows 
us to compute the time-integrated spectrum, starting from the moment when $T_H = 
T_0$.  There is a one-dimensional integral to be done numerically.
\begin{equation}
\frac{dN_{\nu}^{\rm dir}}{dE}=\frac{27m_P^2}{16(4\pi)^6\alpha E^3}
\int_{0}^{E/T_0}dx\frac{x^4}{\exp(x)+1}\left( 0.075+\frac{0.925}
{\exp(5-ax)+1} \right)
\end{equation}
In the high energy limit, meaning $E \gg T_0$, the upper limit can be taken to 
infinity with the result
\begin{equation}
\frac{dN_{\nu}^{\rm dir}}{dE} \rightarrow \frac{29.9m_P^2}{(4\pi)^6\alpha E^3} 
\approx \frac{m_P^2}{565 E^3}\, .
\end{equation}

\subsection{Direct neutrinos from an expanding fluid}

Neutrinos emitted from the decoupling surface have a Fermi distribution in the 
local rest frame of the fluid.  The phase space density is
\begin{equation}
f(E') = \frac{1}{{\rm e}^{E'/T_{\nu}} +1} \, .
\end{equation}
The decoupling temperature of neutrinos is denoted by $T_{\nu}$.  The energy 
appearing here is related to the energy as measured in the rest frame 
of the black hole and to the angle of emission relative to the radial vector by
\begin{equation}
E' = \gamma_{\nu} (1-v_{\nu} \cos\theta) E \, .
\end{equation}
No neutrinos will emerge if the angle is greater than $\pi/2$.  Therefore the 
instantaneous distribution is
\begin{displaymath}
\frac{d^2N_{\nu}^{\rm fluid}}{dE dt} = 
4\pi r_{\nu}^2 \left(\frac{E^2}{2\pi^2}\right)
\int_0^1 d(\cos\theta) \cos\theta f(E,\cos\theta) =
\frac{r^2_{\nu}T_{\nu}}{\pi u_{\nu}} E 
\sum_{n=1}^{\infty}\frac{(-1)^{n+1}}{n}
\end{displaymath}
\begin{equation} 
\left\{ \left( 1 - \frac{T_{\nu}}{n u_{\nu}E} \right)
\exp[-nE (\gamma_{\nu}-u_{\nu})/T_{\nu}]
+ \frac{T_{\nu}}{n u_{\nu}E} \exp[-nE\gamma_{\nu}/T_{\nu}]
\right\} \, ,
\end{equation}
where $r_{\nu}$ is the radius of the decoupling suface and $u = v \gamma$.
Integration over the energy gives the luminosity (per neutrino).

We need to know how the radius and radial flow velocity at neutrino decoupling 
depend on the Hawking temperature or, equivalently, the black hole mass; we have 
already argued that the neutrino temperature at decoupling is about $T_{\nu} = 
100$ GeV.  Our numerical solutions to the viscous fluid equations \cite{us} 
resulted in a simple scaling law between the flow velocity and radius.
\begin{eqnarray}
u(r)=u_S\left(\frac{r}{r_S} \right)^{1/3}
\end{eqnarray}
Here $r_S$ is the Schwarszchild radius and $u_S \approx 0.10$.  (The viscous 
fluid results may not be applicable until the radius exceeds $r_S$ several times 
over.)  Hence $r_{\nu}$ is known if $u_{\nu}$ is.  Specifically
\begin{eqnarray}
r_{\nu}= \frac{1}{4\pi T_H} \left( \frac{u_{\nu}}{u_S} \right)^3 \, .
\end{eqnarray}
The final piece of information is to recognize that each type of neutrino will 
contribute 7/8 effective bosonic degree of freedom to the total number of 
effective bosonic degrees of freedom in the energy density of the fluid.  After 
integrating the instantaneous neutrino energy spectrum to obtain its luminosity 
we equate it with the appropriate fraction of the total luminosity. 
\begin{equation}
L_{\nu}^{\rm fluid}=\frac{7}{8} 
\frac{\pi^3 r_{\nu}^2 T_{\nu}^4}{90} 
\frac{3-v_{\nu}}{(1-v_{\nu})^3 \gamma_{\nu}^4}
= \frac{7/8}{106.75} L^{\rm Hawking}
\end{equation}
where
\begin{equation}
L^{\rm Hawking}= 64\pi^2\alpha T_H^2 \, .
\end{equation}
The number 106.75 counts all effective bosonic degrees of freedom in the 
standard model excepting gravitons.  Without the graviton contribution
$\alpha$ is approximately 0.0042. This results in an equation which determines 
$v$, equivalently $u$, in terms of the Hawking temperature.
\begin{equation}
\frac{(3-v_{\nu})v_{\nu}^4u_{\nu}^2}{(1-v_{\nu})^3} =
2^{13} \frac{45 \pi}{427} \alpha
u_S^6 \left( \frac{T_H}{T_{\nu}}\right)^4 =
C \left( \frac{T_H}{T_{\nu}}\right)^4
\end{equation}
Numerically the constant $C = 1.2\times 10^{-5}$.

We are interested in black holes with $T_H > T_{\nu} = 100$ GeV.  The 
corresponding range of neutrino-sphere flow velocities corresponds to
$0.2 < u_{\nu} < \infty$.  There is no simple analytical expression for the 
solution to eq. (13) for this wide range of the variable.  At asymptotically 
high temperatures the left side has the limit $16 u_{\nu}^8$.  For intermediate 
values the left side may be approximated by $22 u_{\nu}^7$.  We approximate the 
left side by the former for $u_{\nu} > 11/8$ and by the latter for $0.2 < 
u_{\nu} < 11/8$.  Thus       
\begin{eqnarray}
u_{\nu}&=&\left( \frac {CT_H^4}{22T_{\nu}^4}\right)^{1/7} \;\;\; {\rm for}
\;\; u_{\nu} < 11/8 \\
u_{\nu}&=&\left( \frac {CT_H^4}{16T_{\nu}^4}\right)^{1/8} \;\;\; {\rm for}
\;\; u_{\nu} > 11/8 \, .
\end{eqnarray}
This approximation is valid to better than 20\% within the range mentioned.

The time-integrated spectrum can be calculated on the basis of eqs. (8), (3), 
(10) and either (13), which is exact, or with the approximation of (14-15).  The 
resulting formulas are lengthy and not very interesting, therefore not displayed 
here.  What is interesting is the high energy limit, $E \gg 10T_{\nu} = 1$ TeV.  
This is determined wholly by the first of the two exponentials in eq. (8) with 
the coefficient of 1.  Using also eqs. (3), (10) and (14) we find the limit
\begin{equation}
\frac{dN_{\nu}^{\rm fluid}}{dE} \rightarrow
\frac{4185 \zeta(6) C^{1/4}}{854 \pi^5} \frac{m_P^2 T_{\nu}}{E^4}
\approx \frac{m_P^2 T_{\nu}}{1040 E^4}
\end{equation}
This $E^{-4}$ spectrum is characteristic of all neutrino sources in the viscous 
fluid description of the microscopic black hole wind.

\section{Neutrinos from Pion Decay}

The decoupling temperature $T_f$ for pions was estimated in \cite{us} to be in 
the range from 100 to 140 MeV, comparable to the pion mass and substantially 
less than the neutrino decoupling temperature.  Muon-type neutrinos will come 
from the decays of charged pions, namely $\pi^+ \rightarrow \mu^+ \nu_{\mu}$ and 
$\pi^- \rightarrow \mu^- \bar{\nu}_{\mu}$.  In what follows we calculate the 
spectrum of muon neutrinos; the spectrum for muon anti-neutrinos is of course 
identical.

In the rest frame of the decaying pion the muon and the neutrino have momentum q 
determined by energy and momentum conservation.
\begin{equation}
m_{\pi} = q + \sqrt{m_{\mu}^2+q^2}
\end{equation}
Numerically $q = 0.2134 m_{\pi}$.
In the pion rest frame the spectrum for the neutrino, normalized to one, is
\begin{equation}
E^{\prime} \frac{d^3N_{\nu}^{\pi}}{d^3p^{\prime}} =
\frac{\delta(E^{\prime}-q)}{4\pi q} \, .
\end{equation}
The Lorentz invariant rate of emission is obtained by folding together the 
spectrum with the rate of emission of $\pi^+$.
\begin{equation}
E \frac{d^4N_{\nu}^{\pi}}{d^3p dt} = \int_{m_{\pi}}^{\infty} dE_{\pi}
\left( \frac{d^2N_{\pi}}{dE_{\pi}dt} \right) \frac{1}{4\pi q}
\delta \left( \frac{ EE_{\pi} - {\bf p} \cdot {\bf p}_{\pi} }
{m_{\pi}} - q \right)
\end{equation}
The spectrum of pions is computed in the same way as the spectrum of direct 
neutrinos from the expanding fluid in sect. 2.2 but with one difference and one 
simplification.  The difference is that the pion has a Bose distribution as 
opposed to the Fermi distribution of the neutrino.  The simplification is that 
the fluid at pion decoupling has a highly relativistic flow velocity with $u_f 
\approx \gamma_f \gg 1$.  The analog to eq. (8) is
\begin{equation}
\frac{d^2N_{\pi}}{dE dt} = 
- \frac{r_f^2 T_f p_{\pi}}{\pi \gamma_f} \ln \left(1 - 
{\rm e}^{-E_{\pi}/2 \gamma_f T_f} \right) \, .
\end{equation}
The instantaneous energy spectrum of the neutrino is reduced to a single 
integral.
\begin{equation}
\frac{d^2N_{\nu}^{\pi}}{dE dt} = - \frac{m_{\pi} T_f r_f^2}{2\pi q \gamma_f}
\int_{E_{\rm min}}^{\infty} dE_{\pi} 
\ln\left( 1 - {\rm e}^{-E_{\pi}/2\gamma_f T_f} \right)
\end{equation}
Here $E_{\rm min} = m_{\pi}(E^2+q^2)/2Eq$ is the minimum pion energy that will 
produce a neutrino with energy $E$.  We are interested only in high energy 
neutrinos with $E \gg m_{\pi}$, in which case $E_{\rm min} = m_{\pi}E/2q$ is an 
excellent approximation.  The integral can also be expressed as an infinite 
summation
\begin{equation}
\frac{d^2N_{\nu}^{\pi}}{dEdt}=\frac{m_{\pi}r_f^2T_f^2}{\pi q}
\sum_{n=1}^{\infty}\frac{1}{n^2}
\exp\left(-\frac{nm_{\pi}E}{4\gamma_fT_fq}\right)
\end{equation}
We have found from the numerical solutions \cite{us} that
\begin{equation}
\gamma_f T_f = \left( \frac{203 C}{32 d_f} \right)^{1/8}
\sqrt{T_f T_H} \approx 0.22 \sqrt{T_f T_H}
\end{equation}
and that
\begin{equation}
r_f = \frac{1}{\pi u_S^3 T_f} \left( \frac{203 C}{32 d_f} \right)^{3/8}
\sqrt{\frac{T_H}{T_f}} \, .
\end{equation}
Here $d_f = 12$ is the effective number of bosonic degrees of freedom at the 
decoupling temperature $T_f$.  These results take into account the total 
luminosity of the black hole minus that contributed by gravitons and neutrinos.
When $E \gg \gamma_f T_f$ only the first term in the summation in eq. (22) 
is important.

The instantaneous spectrum is easily integrated over time because the flow 
velocity at pion decoupling is highly relativistic.  The spectrum arising from 
the last moments when the Hawking temperature exceeds $T_0$ is
\begin{equation}
\frac{dN_{\nu}^{\pi}}{dE} = \frac{10440}{61 \pi^5 d_f}
\left(\frac{q}{m_{\pi}}\right)^3
\left(\frac{203C}{2d_f}\right)^{1/4}
\frac{m_P^2T_f}{E^4}
\sum_{n=1}^{\infty}\frac{1}{n^2}
\int_0^{m_{\pi}E/4\gamma_f(T_0)T_fq}dx x^3 \, 
{\rm e}^{-nx}
\end{equation}
In the limit that $E \gg \gamma_f(T_0) T_f$ the upper limit on the integral may 
be taken to infinity with the following result.
\begin{equation}
\frac{dN_{\nu}^{\pi}}{dE} \rightarrow \frac{464\pi}{427}
\left(\frac{q}{m_{\pi}}\right)^3
\left(\frac{203C}{2d_f}\right)^{1/4}
\frac{m_P^2T_f}{E^4}
\approx \frac{m_P^2T_f}{300E^4}
\end{equation}
This is much smaller than the direct neutrino emission from the fluid because 
the neutrino decoupling temperature $T_{\nu}$ is much greater than the pion 
decoupling temperature $T_f$.

\section{Neutrinos from Muon Decay}

Muons can be emitted directly or indirectly by the weak decay of pions:
$\pi^- \rightarrow \mu^- \nu_e \bar{\nu}_{\mu}$
plus the charge conjugated decay.  Both sources 
contribute to the neutrino spectrum.  The invariant distribution of the electron 
neutrino (to be specific) in the rest frame of the muon is
\begin{equation}
E^{\prime} \frac{d^3N_{\nu_e}}{d^3p^{\prime}} =
\frac{4}{\pi m_{\mu}^4} (3m_{\mu}-4E^{\prime})E^{\prime}
\end{equation}
where the electron mass has been neglected in comparison to the muon mass.  This 
distribution is used in place of the delta-function distribution of eq. (18) for 
both direct and indirect muons, as evaluated in the following two subsections.

\subsection{Neutrinos from direct muons}

The instantaneous spectrum of $\nu_e$, $\bar{\nu}_e$, $\nu_{\mu}$ or 
$\bar{\nu}_{\mu}$ arising from muons in thermal equilibrium until the decoupling 
temperature of $T_f$ can be computed by folding together the spectrum of muons 
together with the decay spectrum of neutrinos in the same way as eq. (19) was 
obtained.  Using eqs. (20) and (27) results in
\begin{eqnarray}
\frac{d^2N_{\nu}^{{\rm dir} \; \mu}}{dEdt}&=&
\frac{2r_f^2 T_fE}{3\pi\gamma_f}
\sum_{n=1}^{\infty}\Biggl \{ -{\rm Ei}
 \left( -\frac{nE}{2\gamma_fT_f}\right)\left[9\frac{nE}{2\gamma_fT_f}
+2\left(\frac{nE}{2\gamma_fT_f}\right)^2\right] \nonumber\\
&&+\exp\left(-\frac{nE}{2\gamma_fT_f}\right)\left[\frac{10\gamma_fT_f}{nE}-7-
2\frac{nE}{\gamma_fT_f}\right]
\Biggl \},
\end{eqnarray}
where ${\rm Ei}$ is the exponential-integral function.  In the high energy 
limit, defined here by $E \gg 2\gamma_fT_f$, the spectrum simplifies to
\begin{equation}
\frac{d^2N_{\nu}^{{\rm dir} \; \mu}}{dEdt}=\frac{2r_f^2T_fE}{3\pi\gamma_f}
\exp{\left(-\frac{E}{2\gamma_fT_f}\right)} \, .
\end{equation}

The time-integrated spectrum can be calculated in a fashion analogous to that 
followed in section 3.  Thus
\begin{eqnarray}
\lefteqn{\frac{dN_{\nu}^{{\rm dir} \; \mu}}{dE}= 
\frac{870}{61\pi^5 d_f} \left(\frac{203C}{2d_f}\right)^{1/4}
\frac{m_p^2T_f}{E^4}}\nonumber\\ 
&&\times \sum_{n=1}^{\infty} \int_0^{E/2\gamma_f(T_0)T_f} \Biggl\{
-{\rm Ei} \left( -\frac{nE}{2\gamma_fT_f}\right)\left[9\frac{nE}{2\gamma_fT_f}
+2\left(\frac{nE}{2\gamma_fT_f}\right)^2\right] \nonumber\\
&&+\exp\left(-\frac{nE}{2\gamma_fT_f}\right)\left[\frac{10\gamma_fT_f}{nE}-7-
2\frac{nE}{\gamma_fT_f}\right]
\Biggl\} \, . 
\end{eqnarray}
In the high energy limit this simplifies to
\begin{eqnarray}
\frac{dN_{\nu}^{{\rm dir} \; \mu}}{dE} \rightarrow 
\frac{11745}{488\pi^5 d_f} \left[ 225 \zeta(7)-217 \zeta(6) \right]
\left(\frac{203C}{2d_f}\right)^{1/4}
\frac{m_P^2T_f}{E^4} \approx \frac{m_P^2T_f}{250 E^4} \, .
\end{eqnarray}

\subsection{Neutrinos from indirect muons}

The spectrum of neutrinos coming from the decay of muons which themselves came 
from the decay of pions proceeds exactly as in the previous subsection, but with 
the replacement of the direct muon energy spectrum by the indirect muon energy 
spectrum.  In the rest frame of the pion the muon distribution is
\begin{equation}
E^{\prime} \frac{d^3N_{\mu}^{\pi}}{d^3p^{\prime}} =
\frac{\sqrt{m_{\mu}^2+q^2}}{4\pi q^2} \delta(p^{\prime}-q) \, ,
\end{equation}
which is the finite mass version of eq. (18).  Folding together this spectrum 
with the spectrum of pions (20) yields the spectrum of indirect muons.
\begin{equation}
\frac{d^2N_{\mu}^{\pi}}{dEdt}=\frac{m_{\pi}r_f^2T_f^2}{\pi q}
\sum_{n=1}^{\infty}\frac{1}{n^2}
\exp\left(-\frac{nm_{\pi}E}{4\gamma_fT_f\sqrt{m_{\mu}^2+q^2}}\right)
\end{equation}
This spectrum is now folded with the decay distribution of neutrinos to obtain
\begin{eqnarray}
\frac{d^2N_{\nu}^{{\rm indir} \; \mu}}{dEdt} &=&
\frac{m_{\pi}r_f^2 T_f^2}{3\pi q}
\sum_{n=1}^{\infty} \frac{1}{n^2} \Biggl\{
{\rm Ei}(-nx)
\left(5-\frac{27}{6}n^2x^2-\frac{2}{3}n^3x^3\right)
\nonumber\\
&-&\left(-\frac{19}{6}+\frac{23}{6}nx+\frac{2}{3}n^2x^2\right)
{\rm e}^{-nx}
\Biggl\}_{x_-}^{x_+}
\end{eqnarray}
where
\begin{eqnarray}
x_{\pm}=\frac{m_{\pi}E}{2\gamma_fT_f} \left[ 
\frac{(m_{\mu}^2+q^2)^{1/2}\pm q}{m_{\mu}^2}
\right] \, .
\end{eqnarray}
The high energy limit $E \gg \gamma_fT_f$ is
\begin{eqnarray}
\frac{d^2N_{\nu}^{{\rm indir} \; \mu}}{dEdt} \rightarrow
\frac{m_{\pi}r_f^2T_f^2}{3\pi q}
\left[\frac{{\rm e}^{-x_-}}{x_-^2}-\frac{{\rm e}^{-x_+}}{x_+^2}\right] \, .
\end{eqnarray}

The integral over time can be done in the usual way.  The resulting 
expression cannot be expressed in closed form, is lengthy, and is not 
illuminating.  The high energy limit is simple and of the familiar form 
$1/E^4$.  
\begin{equation}
\frac{dN_{\nu}^{{\rm indir} \; \mu}}{dE} \rightarrow 
\frac{174 \pi}{2989 d_f} \left(\frac{203C}{2d_f}\right)^{1/4}
\frac{\sqrt{m_{\mu}^2+q^2} \, (m_{\mu}^2+2q^2)}{m_{\pi}^3}
\frac{m_P^2T_f}{E^4}\nonumber\\
\approx \frac{m_P^2T_f}{1250 E^4}
\end{equation}

\section{Comparison of Neutrino Sources}

In this section we compare the different sources of neutrinos that were computed 
in the previous sections.  All the figures presented display one type of 
neutrino or anti-neutrino.  That type should be clear from the context.  For 
example, equal numbers of $\nu_e$, $\bar{\nu}_e$, $\nu_{\mu}$, 
$\bar{\nu}_{\mu}$, $\nu_{\tau}$, and $\bar{\nu}_{\tau}$ are produced as Hawking 
radiation and by direct emission by the fluid at the neutrino decoupling 
temperature $T_{\nu}$.  Only electron and muon type neutrinos are produced by 
muon decay, and only muon type neutrinos by pion decay.  These differences could 
help to distinguish the decay of a microscopic black hole from other neutrino 
sources if one happened to be within a detectable distance.

The instantaneous spectra are displayed in Fig. 1 for a Hawking temperature of 1 
TeV corresponding to a black hole mass of $10^7$ kg and a lifetime of 7.7 
minutes. The instantaneous spectra for a Hawking temperature of 10 TeV 
corresponding to a black hole mass of $10^6$ kg and a lifetime of 0.5 seconds 
are displayed in Fig. 2.  There are several important features of these spectra.  
One feature is that the spectrum of direct neutrinos emitted by the fluid, at 
the decoupling temperature of $T_{\nu} = 100$ GeV, peaks at a lower energy than 
the spectrum of neutrinos that would be emitted directly as Hawking radiation.  
The peaks are located approximately at $\sqrt{T_{\nu}T_H}$ for the fluid and at 
$5T_H$ for the Hawking neutrinos.  The reason is that the viscous flow degrades 
the average energy of particles composing the fluid, but the number of particles 
is greater as a consequence of energy conservation.  In the viscous fluid 
picture of the black hole explosion direct neutrinos are emitted as Hawking 
radiation without any rescattering when $T_H < T_{\nu}$, whereas when $T_H > 
T_{\nu}$ they are assumed to rescatter and then be emitted from the 
neutrino-sphere located at $T_{\nu}$.  It is incorrect to add the two curves 
shown in these figures.  In reality, of course, it would be better to use 
neutrino transport equations to describe what happens when $T_H \approx 
T_{\nu}$.  That is well beyond the scope of this paper, and perhaps worth doing 
only if and when there is some observational evidence for microscopic black 
holes.

Another feature of Figs. 1 and 2 to note is that the average energy of neutrinos 
arising from pion and muon decay is much less than that of directly emitted 
neutrinos.  Again, the culprit is viscous fluid flow degrading the average 
energy of particles, here the pion and muon, until the time of their decoupling 
at $T_f \approx 100$ MeV.  On the other hand their number is greatly increased 
on account of energy conservation.  The average energies of the neutrinos are 
somewhat less than their parent pions and muons because energy must be shared 
among the decay products.  The spectrum of muon-neutrinos coming from the decay
$\pi \rightarrow \mu \nu$ is the softest because the pion and muon masses
are very close, leaving very little energy for the neutrino.

The time-integrated spectra, starting at the moment when the Hawking temperature 
is 1 and 10 TeV, are shown in Figs. 3 and 4, respectively.  The relative 
magnitudes and average energies reflect the trends seen in Figs. 1 and 2.  At 
high energy the Hawking spectrum is proportional to $E^{-3}$ while all the 
others are proportional to $E^{-4}$, as was already pointed out in the previous 
sections.  Obviously the greatest number of neutrinos by far are emitted at 
energies less than 100 GeV.  The basic reason is that only about 5\% of the 
total luminosity of the black hole is emitted directly as neutrinos.  About 32\% 
goes into neutrinos coming from pion and muon decay, about 24\% goes into 
photons, with most of the remainder going into electrons and positrons.

\section{Observability of the Neutrino Flux}

We now turn to the possibility of observing neutrinos from a microscopic black 
hole directly.  Obviously this depends on a number of factors, such as the 
distance to the black hole, the size of the neutrino detector, the efficiency of 
detecting neutrinos as a function of neutrino type and energy, how long the 
detector looks at the black hole before it is gone, and so on.

For the sake of discussion, let us assume that one is interested in neutrinos 
with energy greater than 10 GeV and that the observational time is the last 7.7 
minutes of the black hole's existence when its Hawking temperature is 1 TeV and 
above.  As may be seen from Fig. 3, most of the neutrinos will come from the 
decay of directly emitted muons.  Integration of eq. (31) from $E_{\rm min} = 
10$ GeV to infinity, and multiplying by 4 to account for both electron and muon 
type neutrinos and anti-neutrinos, results in the total number of
\begin{equation}
N_{\nu} =\frac{4}{750} \frac{m_P^2 T_f}{E_{\rm min}^3}
\approx 8\times 10^{31} \, .
\end{equation}
This does not take into account neutrinos directly emitted from the fluid.  For   
$E_{\rm min} = 1$ TeV, for example, eq. (16) should be used in place of eq. (31) 
(see Fig. 3), and taking into account tau-type neutrinos too then yields a total 
number of about $3 \times 10^{28}$.  For an exploding black hole located a distance $d$ 
from Earth the number of neutrinos per unit area is
\begin{eqnarray}
N_{\nu}(E > 10\;{\rm GeV}) &=& 
6700 \left(\frac{1\;{\rm pc}}{d}\right)^2 \; {\rm km}^{-2} \, ,\\
N_{\nu}(E > 1\;{\rm TeV}) &=& 
2.5 \left(\frac{1\;{\rm pc}}{d}\right)^2 \; {\rm km}^{-2} \, .
\end{eqnarray}
Although the latter luminosity is smaller by three orders of magnitude, it has 
two advantages.  First, 1/3 of that luminosity comes from tau-type neutrinos.  
Unlike electron and muon-type neutrinos, the tau-type is not produced by the 
decays of pions produced by interactions of high energy cosmic rays with matter 
or with the microwave background radiation.  Hence it would seem to be a much 
more characteristic signal of exploding black holes than any other cosmic source 
(assuming no oscillations between the tau-type and the other two species).  
Second, the tau-type neutrinos come from near the neutrino-sphere, thus probing 
physics at a temperature of order 100 GeV much more directly than the other 
types of neutrinos.

What is the local rate density $\dot{\rho}_{\rm local}$ of exploding black 
holes?  This is, of course, unknown since no one has ever knowingly 
observed a black hole explosion.  The first observational limit was determined 
by Page and Hawking \cite{PH}.  They found that the local rate density
is less than 1 to 10 per cubic parsec per year on the 
basis of diffuse gamma rays with energies on the order of 100 MeV.  This limit 
has not been lowered very much during the intervening twenty-five years.  For 
example, Wright \cite{W} used EGRET data to search for an anisotropic
high-lattitude component of diffuse gamma rays in the energy range from 30 MeV 
to 100 GeV as a signal for steady emission of microscopic black holes.  He 
concluded that $\dot{\rho}_{\rm local}$ is less than about 0.4 per cubic parsec 
per year.  If the actual rate density is anything close to these upper limits 
the frequency of a high energy neutrino detector seeing a black hole explosion 
ought to be around one per year.

\section{Conclusion}

This paper has been a continuation of our previous work on a viscous fluid 
description of the radiation from microscopic black holes.  In this paper we 
have calculated the spectra of all three flavors of neutrinos arising from 
direct emission from the fluid at the neutrino-sphere and from the decay of 
pions and muons from their decoupling at much larger radii and smaller 
temperatures.  We should emphasize in particular the usefulness of 
distinguishing between the electron and muon-type neutrinos and the tau-type 
ones.  The latter are much less likely to be produced by high energy cosmic 
rays.  If the rate density of exploding microscopic black holes in our vicinity 
is anywhere close to the current limit based on gamma rays, it should be 
possible to observe them with present and planned large astrophysical neutrino 
detectors.  Observation of high energy neutrinos, especially in conjunction with 
high energy gamma rays, may provide a window on physics well beyond the TeV 
scale.

\section*{Acknowledgements}

We are grateful to Y.-Z. Qian for comments on the manuscript.  This work was 
supported by the US Department of Energy under grant DE-FG02-87ER40328.

\newpage

\begin{figure}
\centerline{\epsfig{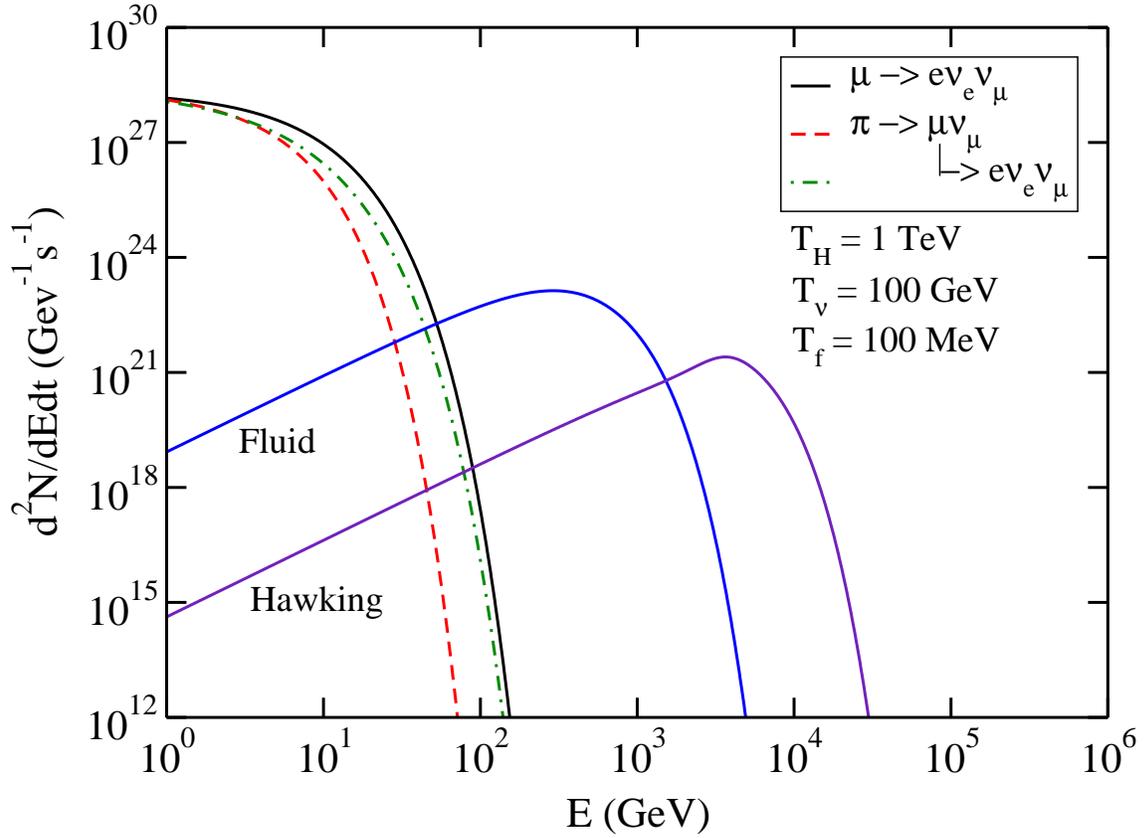}}
\caption{The instantaneous direct neutrino spectra emerging from the  
fluid with neutrino-sphere located at $T_{\nu}=100$ GeV compared to the direct 
Hawking radiation.  Also shown are neutrinos arising from  
pion, direct muon and indirect muon decays  at a decoupling temperature of 
$T_f=100$ MeV.  Here the black hole temperature is $T_H=1$ TeV.
All curves are for one flavor of neutrino or anti-neutrino.}
\end{figure}

\begin{figure}
\centerline{\epsfig{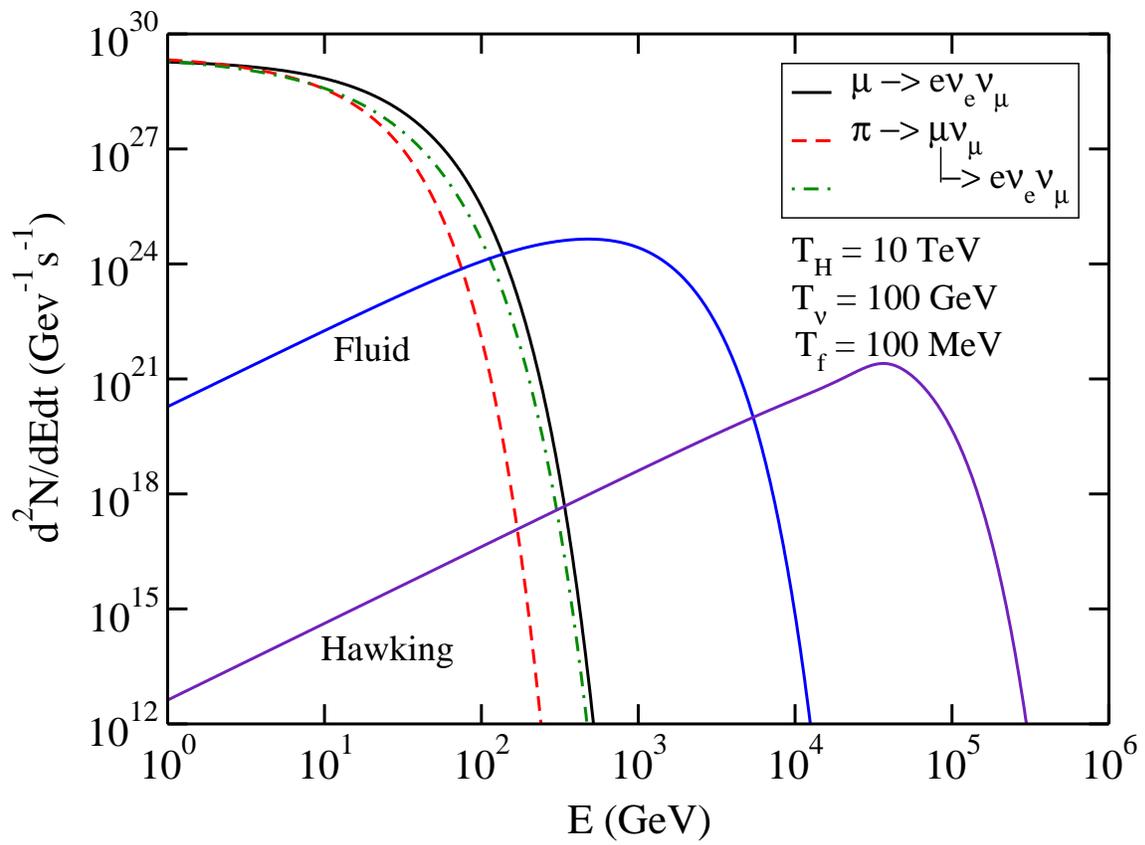}}
\caption{Same as Fig. 1 but with $T_H=1$ TeV.}
\end{figure}

\begin{figure}
\centerline{\epsfig{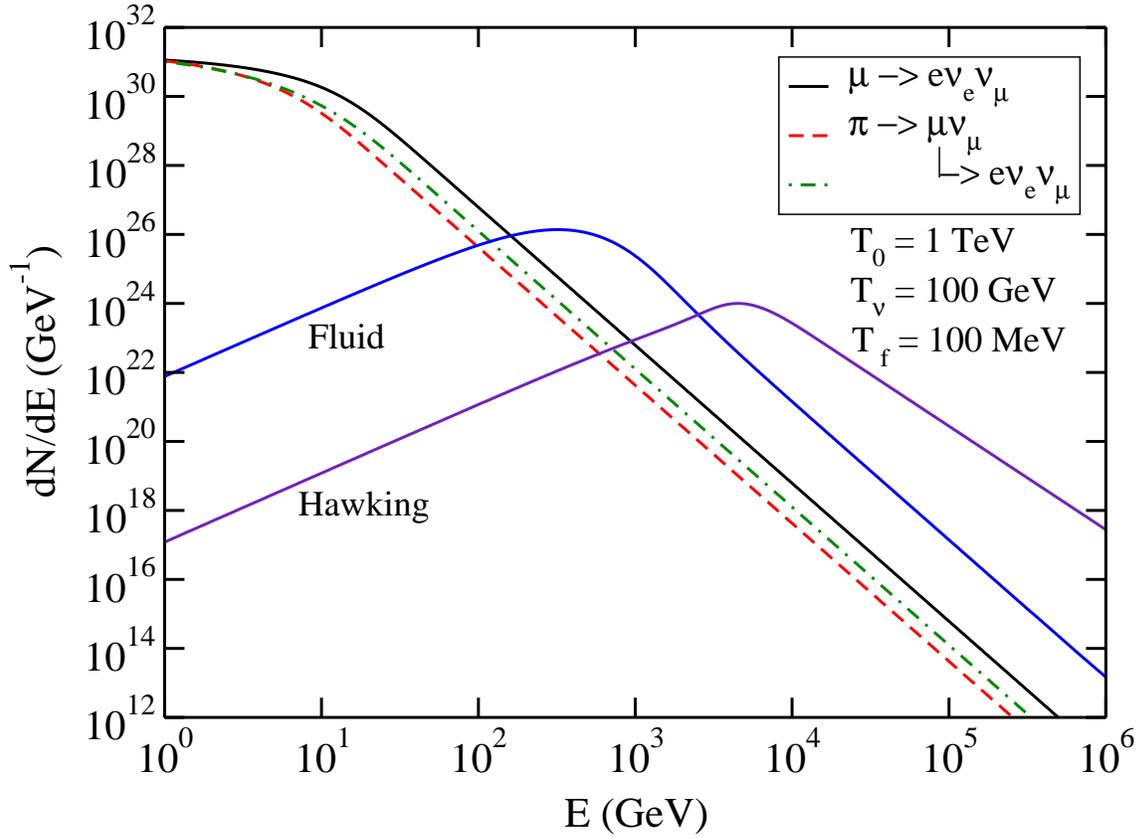}}
\caption{The time integrated neutrino spectra emerging from  
a microscopic black hole.  Here the calculation begins when the black hole 
temperature is $T_0=1$ TeV.  Either the direct Hawking radiation of neutrinos 
or the direct neutrino emission from a neutrino-sphere at a 
temperature of 100 GeV should be used.  All curves are for one species of 
neutrino or anti-neutrino.}
\end{figure}

\begin{figure}
\centerline{\epsfig{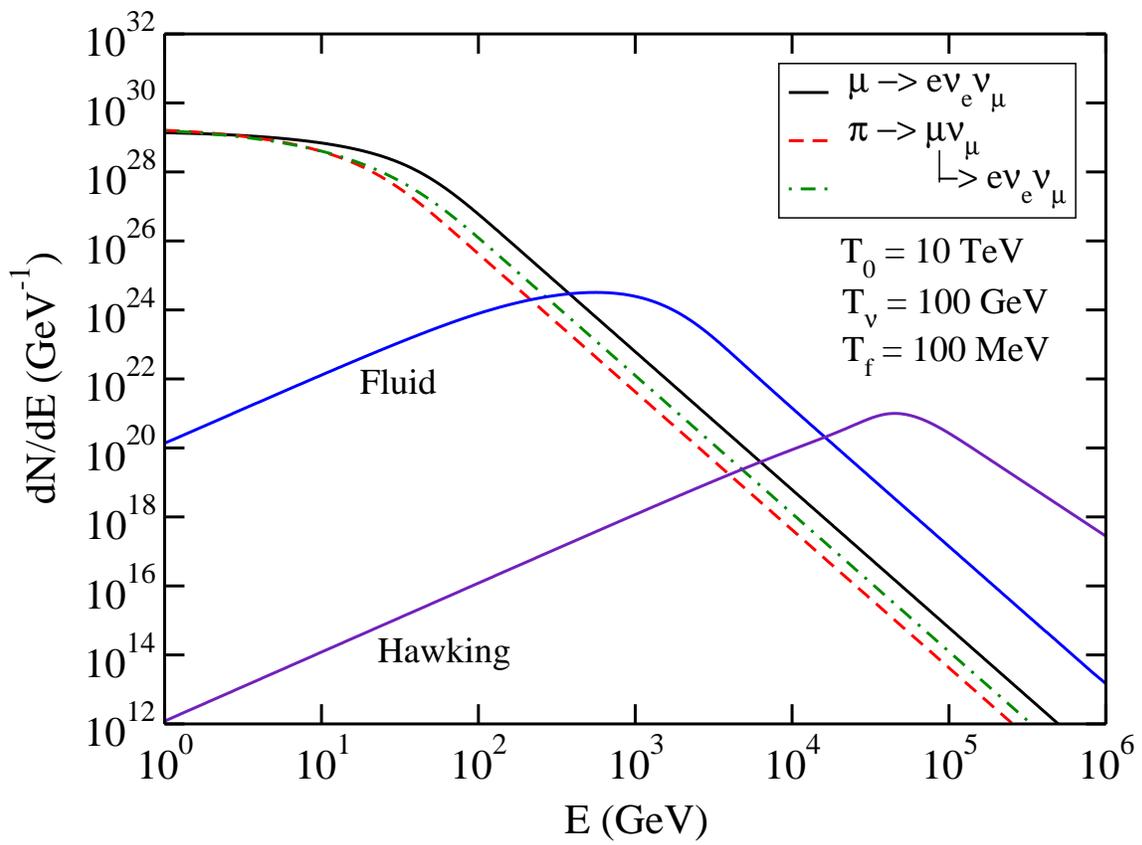}}
\caption{Same as Fig. 3 but with $T_0=10$ TeV.}
\end{figure}

\end{document}